\documentclass[epj]{svjour}

\usepackage{graphics}
\usepackage{float}
\usepackage{bm}
\usepackage{comment}
\usepackage{soul}
\usepackage{cite} 
\usepackage{hyperref}
\usepackage{caption} 
\usepackage{threeparttable}
\usepackage{amssymb}
\usepackage{amsmath}

\begin{document}

\title{New limit for the half-life of double beta decay of $^{94}$Zr to the first excited state of $^{94}$Mo}

\author{N.~Dokania\inst{1}, V.~Nanal\inst{1}\thanks{\email{nanal@tifr.res.in}}, G.~Gupta\inst{1}, S.~Pal\inst{2}, R.~G.~Pillay\inst{1}, P.~K.~Rath\inst{3}, V.I.~Tretyak\inst{4, 5}, A.~Garai\inst{6, 7}, H.~Krishnamoorthy\inst{6, 7}, C.~Ghosh\inst{1}, P.~K.~Raina\inst{8}\and K.~G.~Bhushan\inst{9}
}                     
\institute{Department of Nuclear and Atomic Physics, Tata Institute of Fundamental Research, Mumbai 400 005, India \and Pelletron Linac Facility, Tata Institute of Fundamental Research, Mumbai 400 005, India \and Department of Physics, University of Lucknow, Lucknow 226007, India \and Institute for Nuclear Research, MSP 03680 Kyiv, Ukraine \and INFN, sezione di Roma, I-00185 Rome, Italy \and India based Neutrino Observatory, Tata Institute of Fundamental Research, Mumbai 400 005, India \and Homi Bhabha National Institute, Anushaktinagar, Mumbai 400 094, India \and Department of Physics, Indian Institute of Technology, Ropar, Rupnagar 140 001, India \and Technical Physics Division, Bhabha Atomic Research Centre, Mumbai 400 085, India}
\date{Received: date / Revised version: date}
%
\abstract{Neutrinoless Double Beta Decay is a phenomenon of fundamental interest in particle physics. The decay rates of double beta decay transitions to the excited states can provide input for Nuclear Transition Matrix Element calculations for the relevant two neutrino double beta decay process. It can be useful as supplementary information for the calculation of Nuclear Transition Matrix Element for the neutrinoless double beta decay process. In the present work, double beta decay of $^{94}$Zr to the $2^{+}_{1}$ excited state of $^{94}$Mo at 871.1 keV is studied using a low background $\sim$ 230 cm$^3$ HPGe detector. 
No evidence of this decay was found with a 232 g.y exposure of natural Zirconium.
 The lower half-life limit obtained for the double beta decay of $\rm^{94}Zr$ to the $2^{+}_{1}$ excited state of $\rm^{94}Mo$ is $T_{1/2} (0\nu + 2\nu)> 3.4 \times 10^{19}$ y at 90\% C.L., an improvement by a factor of $\sim$ 4 over the existing experimental limit at 90\% C.L. The sensitivity is estimated to be $T_{1/2} (0\nu + 2\nu) > 2.0\times10^{19}$ y at 90\% C.L. using the Feldman-Cousins method.
\PACS{
      {23.40.-s}{double beta decay}   \and
      {07.85.Fv}{gamma ray detectors}
     } 
} 
\hugehead
\maketitle
\hugehead
\section{Introduction}

\label{intro}
Double beta decay (DBD) is a rare second-order weak nuclear transition, first suggested by Maria Goeppert-Mayer in 1935~\cite{mayer}. Generally, in the case of a beta unstable parent nucleus it is extremely difficult to distinguish the rare DBD from an intensive single beta decay background. There are 35 even-even nuclides with single beta decay either energy forbidden or spin suppressed, which makes it possible to search for the DBD transformation in these nuclides. In the two-neutrino double beta decay ($2\nu\beta\beta$) mode, two neutrons simultaneously undergo beta decay producing two protons, two electrons and two anti-neutrinos in the final state. The $2\nu\beta\beta$ process has been experimentally observed in 12 nuclei so far with a half-life range of $T_{1/2} \sim 10^{18} - 10^{24}$ y~\cite{saakyan}. For a number of nuclei, double beta decays to excited states in their daughter nuclei are also energetically possible. These processes can be probed at low background facilities and limits of the order of $T_{1/2} \sim 10^{18} - 10^{25}$ y on different nuclei have been reached using different experimental techniques~\cite{tretyak, 124Sn, 130Te, 100Mo, 76Ge, 82Se, 96Zr, 110Pd, 116Cd, asakura}.

The most interesting mode is neutrinoless double beta decay ($0\nu\beta\beta$)~\cite{furry} where the neutrinos are not emitted in the final state and which is predicted to occur in extensions of the Standard Model of particle physics. Observation of $0\nu\beta\beta$ would provide evidence for lepton number violation, Majorana nature of neutrinos and can yield measurement of the effective neutrino mass. 
The important implications of $0\nu\beta\beta$ observations~\cite{pas, vergados} and demonstration of neutrino oscillation experiments~\cite{nuosc} have triggered a new generation of $0\nu\beta\beta$ experiments using a variety of candidate isotopes and experimental techniques~\cite{cremonesi, stefano, bilenky}. 
For the $0\nu\beta\beta$ process mediated via the exchange of a light mass Majorana neutrino, the decay rate is related to the effective Majorana neutrino mass through the Nuclear Transition Matrix Element (NTME). The NTME calculations are model-dependent and have large ``theoretical'' uncertainties, limiting the sensitivity of the neutrino mass extraction from the observation of $0\nu\beta\beta$~\cite{vergados2}.
The measurement of DBD transitions to the excited states can constrain the nuclear models used for the calculations of NTME for the 2$\nu\beta\beta$ ($M^{2\nu}$). This experimental input can also provide supplementary information on the NTME calculations for the 0$\nu\beta\beta$ ($M^{0\nu}$)~\cite{jouni}, thereby reducing the uncertainty in the neutrino mass. Further, the difference between the decays to the different states of the final nuclei can be used to constrain/study the other possible mechanisms of $0\nu\beta\beta$ such as exchange of right-handed W-bosons or of supersymmetry (SUSY) models with R-parity violation~\cite{susy, simkovic}.

There are many ongoing efforts to search for the DBD processes in different isotopes using a wide variety of experimental techniques. The DBD to excited states is generally studied by measurement of discrete characteristic gamma rays in the daughter nucleus. Positive signals are seen in the cases of $^{100}$Mo~\cite{100Mo_1995, 100Mo_1999, 100Mo_2007, 100Mo_2009, 100Mo_2010, braeck, 100Mo} and $^{150}$Nd~\cite{150Nd_2004, 150Nd_2009, 150Nd_2014} while in other cases lower half-life limits are obtained for the decay process~\cite{saakyan}. Moreover, measurement of such rare processes is an experimental challenge in itself. The understanding and reduction of background is a crucial factor in improving the sensitivity of the half-life measurement for DBD. Ultra-low levels of background are required to reach the desired sensitivity~\cite{agositini, auger}.

\begin{figure}[H]
\centering
\includegraphics{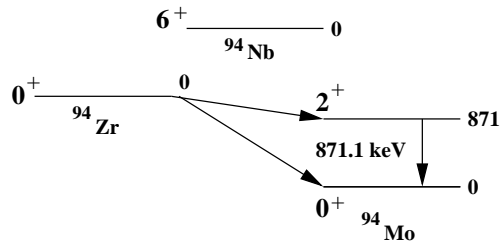}
\caption{\label{fig:zr_decay} Decay scheme of $^{94}$Zr.}
\end{figure}

The TiLES (Tifr Low background Experimental Setup) comprising a high efficiency HPGe detector has been set up at sea level at TIFR, Mumbai~\cite{nima} and is used to study DBD in $^{94}$Zr to the excited state of its daughter nucleus. There are two DBD isotopes of Zr, namely $^{94}$Zr and $^{96}$Zr. The normal beta decay in $^{96}$Zr is spin suppressed while that for $^{94}$Zr is energetically forbidden. The DBD in $^{96}$Zr has been studied~\cite{arpesella, arnold, 96Zr, nemo} due to its higher Q$_{\beta\beta}$ value (3356.097 (86) keV~\cite{alanssari}) and recent results put the half-life limit at $T_{1/2}>3.1\times10^{20}~\rm y~(^{96}\rm{Zr}\rightarrow~^{96}\rm{Mo}, 0^{+}\rightarrow0^{+}_{1})$~\cite{nemo}. The $^{94}$Zr isotope has a lower Q$_{\beta\beta}$ value (1144.56 (31) keV~\cite{gulyuz}) but has a high natural abundance ($\sim17$\%). Fig.~\ref{fig:zr_decay} shows the decay scheme for $^{94}$Zr where the single beta decay to $^{94}$Nb is kinematically forbidden while DBD to either 2$^+_1$ (excited state) or 0$^+$ (g.s.) of $^{94}$Mo is allowed. There has been only one reported study of the DBD of $^{94}$Zr to the 2$^{+}_{1}$ excited state of $^{94}$Mo by Norman {\it et al.}~\cite{norman}, where they have given a lower limit as $T_{1/2} > 1.3\times10^{19}$ y at 68\% C.L. with $\sim$ 115 g.y exposure (646 g of $\rm^{nat}$Zr, 65 days of data). This corresponds to a limit of $T_{1/2} > 0.8\times10^{19}$ y at 90\% C.L. .
Recent theoretical calculations by J. Suhonen using the Quasi-particle Random Phase Approximation (QRPA) model for NTME calculation gave $T_{1/2}~(^{94}\rm{Zr}\rightarrow~^{94}\rm{Mo}, 0^{+}\rightarrow2^{+}_{1})$ to be $\sim10^{32}$ y for 2$\nu\beta\beta$ mode~\cite{suho11}. Due to the large uncertainties involved in the NTME calculations and the theoretical approaches used for the predictions of half-life, experimental data of $T_{1/2}$ with improved sensitivity in different nuclei are highly essential. The motivation for the present work is to search for DBD of $^{94}$Zr to the 2$^{+}_{1}$ excited state of $^{94}$Mo at 871.1 keV with better sensitivity. The $^{94}$Zr study was also done keeping in mind that double beta decay with $\pm2$ neutron numbers have similar NTME values~\cite{be2}. Hence, it could be also relevant for the $^{96}$Zr isotope which is more favoured due to its high Q$_{\beta\beta}$ value. As mentioned earlier, it is an experimental challenge to study DBD to excited states and there is only one existing measurement for $^{94}$Zr.
The lower half-life limit obtained in the present work is $T_{1/2} (0\nu + 2\nu) > 3.4 \times 10^{19}$ y at 90\% C.L.  using a 232 g.y exposure of natural Zirconium. The new half-life limit is improved by a factor of $\sim 4$ over the existing experimental limit. Experimental details are discussed in Section 2. Data analysis and results of the measurement are presented in Section 3 while the summary is given in Section 4.

\section{Experimental Details}

The TiLES consists of a low background, high efficiency HPGe detector (70\% relative efficiency). The detector is a p-type HPGe with an active volume $\sim$ 230 cm$^3$, biased to +4 kV and top Ge dead layer is 1.2 mm with a 0.9 mm carbon fiber endcap. Typical energy resolution measured is FWHM $\sim$ 2.2 keV at 1332.2 keV.
The detector is shielded with 5 cm low activity OFHC Cu, 10 cm low activity Pb ($\rm^{210} Pb < 0.3$ Bq/kg), a Radon exclusion box and an active muon veto system using three plastic scintillators (P1, P2, P3). The box is an air-tight 6 mm thick Perspex box surrounding the HPGe detector as well as the Pb+Cu shield. This box volume is continuously purged with boil-off N$_2$ at an over-pressure of $\sim8$ mbar to reduce the Radon ($^{222}$Rn) contamination. 
\begin{figure}[H]
\centering
\resizebox{0.4\textwidth}{!}{\includegraphics{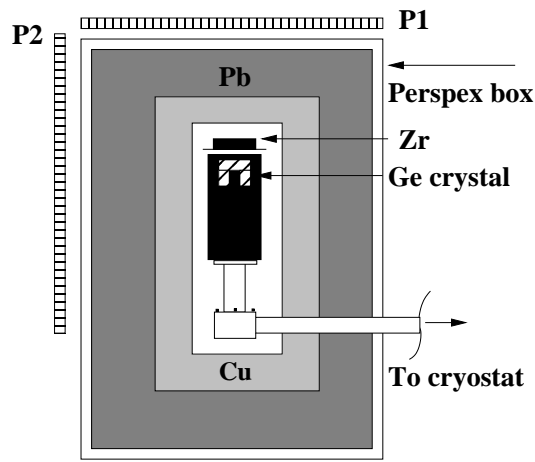}}
\caption{\label{fig:tiles} A picture of TiLES with the shielding arrangement and the mounting of the Zr sample on the Perspex plate holder. See text for details. The third plastic scintillator P3 is not shown here.}
\end{figure}
The three plastic scintillators (50 cm $\times$ 50 cm $\times$ 1 cm each) are arranged outside the Perspex box in a geometry to obtain the best possible muon coverage ($\sim60$\%). Fig.~\ref{fig:tiles} shows a schematic view of the HPGe detector and the surrounding shielding arrangement (third plastic scintillator P3 is not shown).
The response of the HPGe detector has been completely characterized with Geant4-based Monte Carlo (MC) simulations~\cite{nima}. This allows the extraction of the detection efficiency with an accuracy of 5\% for different source configurations counted in a close geometry required in low background gamma spectroscopy. 
The pulses from the preamplifier of the detector are fed to a 14-bit, 100 MHz commercial CAEN-based N6724 digitizer. The digitizer produces the time stamp and the energy deposited in the detector on an event-by-event mode. The algorithm implemented for pulse height analysis is based on the trapezoidal filter (moving window de-convolution). 
The unit employs a Timing and Trigger Filter which performs a digital RC-CR$^2$ filter, whose zero crossing corresponds to the trigger time stamp, i.e., the time of arrival of the pulse. The plastic scintillator signals (rise time $\sim$ 8 ns) are given through a fast amplifier to stretch the signal rise times in order to make it compatible with the sampling rate of the digitizer (100 MHz). An offline algorithm is used to perform the anti-coincidence between the Ge detector and the plastic scintillator signals. The coincidence window is defined to be $\pm2.5~\mu s$. The data from the digitizer is analyzed using the ROOT analysis framework~\cite{root} and LAMPS software~\cite{lamps}. The stability of the energy scale is monitored with background gamma rays such as 661.7, 1460.8 and 2614.5 keV and is found to be better than 1\% over a period of 12 weeks. It should be mentioned that the energy calibration using $^{152}$Eu gamma source was done at the start and the end of the Zr counting. In addition, a standard 10 Hz pulser input given to the charge sensitive preamplifier is used for continuous monitoring. This was also used to estimate the live time of the counting setup, which is found to be 99.5\%.

\begin{figure}[H]
\begin{center}
\resizebox{0.28\textwidth}{!}{\includegraphics{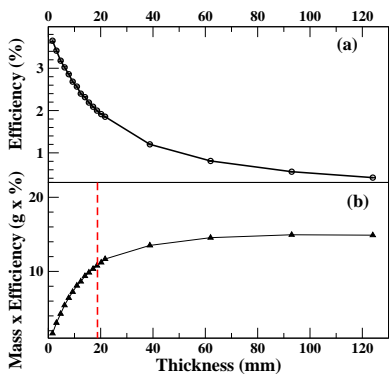}}
\caption{\label{fig:zreff}(Color online) (a) Comparison of efficiency of 871.1 keV gamma ray and (b) mass times efficiency for different thickness of the Zr sample. Lines are only a guide to the eye. }
\end{center}
\end{figure}

Natural Zirconium in the form of 1.55 mm thick plates (99.9\% purity, Princeton Scientific Corp.) was mounted in a close geometry in the TiLES. It should be mentioned that the Zr sample was procured in form of bigger plates (60 mm x 300 mm, 1.55 mm thick) from the same set and then cut to desired size for the counting. The composition analysis and impurity assessment of the sample was done using SIMS (Secondary Ion Mass Spectrometry)~\cite{sims} and ICPMS (Inductively-Coupled Plasma Mass Spectrometry)~\cite{icpms}. The abundance of $^{94}$Zr was measured to be 16.58(17)\% in the sample, a deviation of about 5\% from the isotopic natural abundance 17.38(4)\%~\cite{nndc}. Two different samples of 10 mm x 10 mm were sent for both ICPMS and SIMS analyses and showed identical composition within the errors.  
Table~\ref{tab:sims} gives the concentration of the impurity elements present in the Zr sample obtained from the SIMS analysis. 
The ICPMS measurements of the Zr sample gave 200 ppb and 25 ppb of $^{238}$U and $^{232}$Th respectively and $^{235}$U is estimated to be $<$ 5 ppb. The uncertainties on the SIMS and ICPMS results are within 1\%.
\begin{table}
\centering
\caption{Impurity levels in the Zr sample obtained from the SIMS measurement.}
\label{tab:sims}
\begin{tabular}{  ll }
\hline\noalign{\smallskip}
Element & Concentration \\
        &       (ppm)   \\
\noalign{\smallskip}\hline\noalign{\smallskip}
Hf      &        135    \\
Fe      &        55    \\
Al      &        40    \\
Ni      &        15    \\
Cr      &        12    \\
Sn      &        12    \\
All other elements      &        $<$10    \\
\noalign{\smallskip}\hline
\end{tabular}
\end{table}
The thickness and hence mass of the Zr sample to be mounted was optimized using MC simulations. In the present work, GEANT4 (version 4.9.5.p01) \cite{geant} is used for the MC simulations. The mass attenuation of the 871.1 keV gamma ray was taken into account and the results are shown in fig.~\ref{fig:zreff}. It is evident from fig.~\ref{fig:zreff}(b) that the product of the mass and efficiency of 871.1 keV gamma ray saturates beyond 40 mm thick Zr sample. Given the material availability, $\sim$ 20 mm thick Zr sample was used and is shown by the dotted red line in fig.~\ref{fig:zreff}(b). 
\vspace{1.5 cm}
The 1.55 mm thick Zr plates (60 $\times$ 75 mm$^2$) were stacked to make the 18.6 mm thick Zr block. It was mounted on a 2 mm thick Perspex plate kept at a distance of 7 mm from the top face of the detector (see fig.~\ref{fig:tiles}). The counting was done in stages with Zr mass of 180 g (60 $\times$ 75 $\times$ 6.2 mm$^3$), 360 g (60 $\times$ 75 $\times$ 12.4 mm$^3$), 540 g (60 $\times$ 75 $\times$ 18.6 mm$^3$) and the background in the region of interest (ROI, $E_\gamma=$ 820-920 keV) was monitored at each stage. For the 180 g and 360 g data, the shield configuration was Pb+Cu+P1+P2. Before mounting the 540 g Zr plates, the shielding setup was upgraded with P3 + N2 flushing, which led to the reduction of background by $\sim$10\% in the ROI. The background in either shielding configuration was measured with the Perspex sample holder. The detection efficiency at $E_{\gamma}=$ 871.1 keV was determined using the MC detector model employing the actual geometry and mounting setup in the simulations. The self-absorption of the 871.1 keV gamma ray in the Zr block has been taken into account in the MC simulations and is found to be about 32\% for the 540 g Zr sample.
The energy resolution (FWHM) at $E_{\gamma}=$ 871.1 keV is about 2.10 $\pm$ 0.01 keV, obtained from the fit to the resolution curve using $^{152}$Eu. The energy resolution has been also measured for the 834.8 keV gamma ray using $^{54}$Mn source and the measured FWHM was 2.06 $\pm$ 0.01 keV. The integral background rate achieved for the energy region 40 -- 2700 keV divided by the mass of the Ge crystal is 2$\times10^4$ /day/kg, better than the levels achieved in some of the HPGe-based surface laboratories~\cite{laubenstein}. The background level quoted is for the complete shielding setup without Zr, namely, HPGe + Pb + Cu + P1 + P2 + P3 + N2 flushing in a Perspex box. It should be mentioned that the background level at the surface is limited by the cosmic-muon induced interactions in the high-$Z$ Pb and Cu shield materials. Therefore, the background level achieved in the present setup with veto for cosmic muon rejection is better than that reported in Ref.~\cite{laubenstein} with passive shielding only.

\section{Data Analysis and Results}
The data set considered for the analysis was taken with 180 g, 360 g and 540 g amounting to a total exposure of 232 g.y of natural Zirconium.
\vspace{1.8 cm}
\begin{figure}[H]
\begin{center}
\resizebox{0.8\textwidth}{!}{\includegraphics{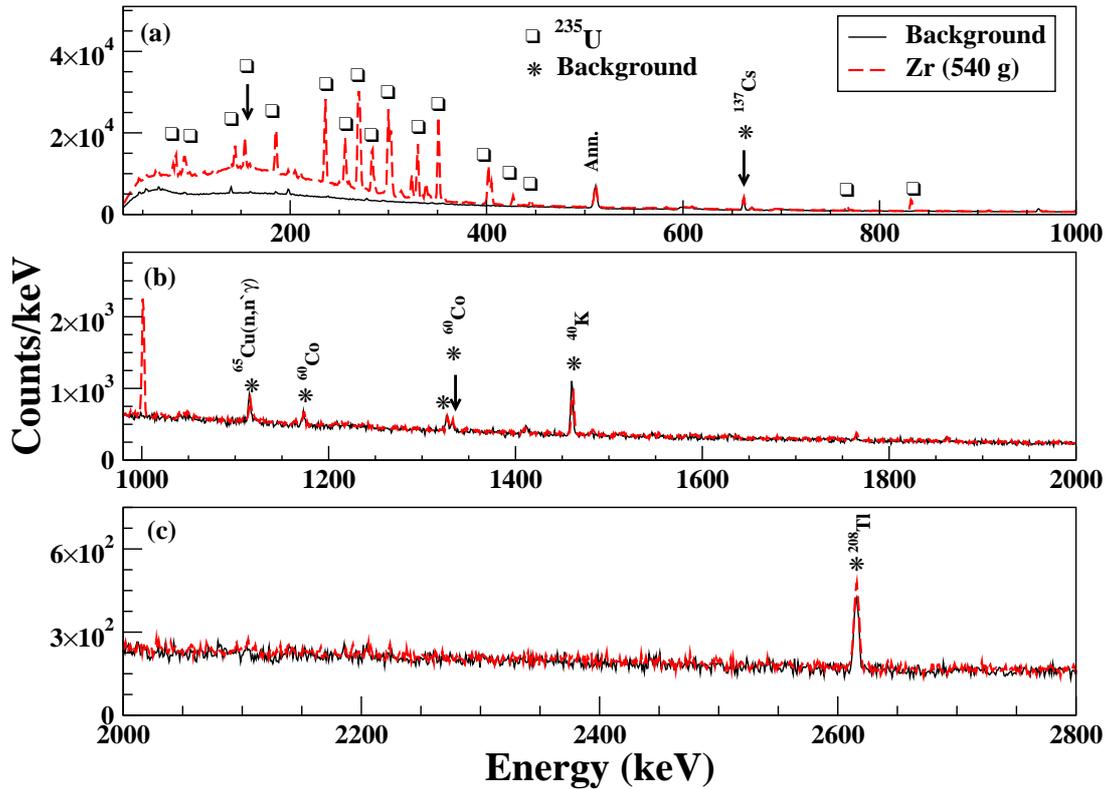}}
\caption{\label{fig:bkg} (Color online) Energy spectra of the room background (solid black line) together with 540 g Zr sample (dotted red line) for a counting time of 75 days (see text for details) .}
\end{center}
\end{figure}
Due to the low counting rate, there are negligible pile-up events in the data. The room background data was collected for 56 days and 75 days with the shielding configurations I (Pb+Cu+P1+P2) and II (Pb+Cu+P1+P2+P3+N$_2$), respectively. Fig.~\ref{fig:bkg} shows the energy spectrum of the 540 g Zr sample counted for a period of 75 days together with the room background spectrum corresponding to shielding configuration II. In the TiLES, the background after shielding arises mainly from $^{40}$K (primordial) and impurities such as $^{60}$Co, $^{137}$Cs. In addition, there are some gamma rays observed due to the neutron interactions in Ge~\cite{harisree}.
There are many gamma rays seen above the background level in the energy spectrum with the Zr sample. 
Most of the gamma rays are produced due to the isotopes in the decay chain of $^{235}$U, present as a trace impurity in the Zr sample. 
\begin{figure}
 \centering
\resizebox{0.6\textwidth}{!}{\includegraphics{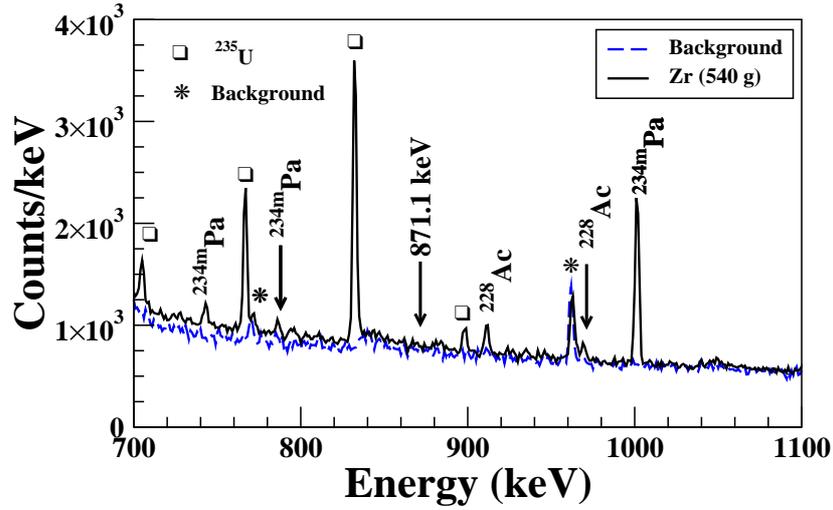}}
\caption{\label{fig:bkg_enl} (Color online) Expanded region ($E_{\gamma}$= 700-1100 keV) of the energy spectrum of the 540 g Zr sample (solid black) together with the room background spectrum (dotted blue) for a counting time of 75 days (see text for details). }
\end{figure}
Fig.~\ref{fig:bkg_enl} shows the energy spectra in the enlarged energy window of 700-1100 keV of the 540 g Zr sample together with the room background corresponding to shielding configuration II for a counting period of 75 days.
Table~\ref{tab:gammalist} gives the list of major gamma rays ($E_{\gamma}>100$ keV) observed in the Zr sample above the room background level and their source of origin.
\begin{table}[H]
\centering
\caption{List of the major gamma rays ($E_{\gamma}>100$ keV) together with their source of origin as observed in the Zr sample above the room background level. }
\label{tab:gammalist}
\begin{tabular}{  llll }
\hline\noalign{\smallskip}

 $E_{\gamma}$	& Source  & 	$E_{\gamma}$	&  		Source	         \\
  (keV)     &      	     & 		(keV)      &       	\\
  \noalign{\smallskip}\hline\noalign{\smallskip}
  113.1     &   $^{227}$Th	& 	329.8, 330.1& $^{227}$Th, $^{231}$Pa	\\
  122.3&$^{223}$Ra	& 		334.4		& 	$^{227}$Th			\\
  143.8, 144.2&$^{235}$U, $^{223}$Ra	&	338.3	& $^{223}$Ra\\
 154.2 & $^{223}$Ra &  351.1	  &  $^{211}$Bi \\ 
   158.6  & $^{223}$Ra	& 		371.7	& 	$^{223}$Ra		\\
 163.3   & $^{235}$U		& 	401.8		& 	$^{219}$Rn		\\
 185.7, 186.2  & $^{235}$U,  $^{226}$Ra 	& 404.9  & $^{211}$Pb	\\
  205.1 & $^{235}$U	&   427.1	   	& 	$^{211}$Pb	\\ 
  210.6   &  $^{227}$Th	& 	445.0	& 	 $^{223}$Ra			\\ 
235.9  & $^{227}$Th	& 		583.2		& 	$^{208}$Tl			\\ 
 249.3 & $^{223}$Ra	& 			742.8	& 	$^{234m}$Pa		\\ 
  256.2, 255.8   & $^{227}$Th, $^{231}$Pa	& 766.5, 766.4& $^{211}$Pb, $^{234m}$Pa	\\	
 260.2 &	$^{231}$Pa  & 	 	786.3	& 		$^{234m}$Pa	\\ 
 269.5, 271.2 &	$^{223}$Ra, $^{219}$Rn & 	 	832.0	& $^{211}$Pb\\	
 277.2  &	$^{231}$Pa  & 	 	897.8		& 		$^{207}$Tl		\\	
  283.7  &	$^{231}$Pa  & 	911.2		& 	$^{228}$Ac	\\ 	
 299.9, 300.1 &	 $^{227}$Th, $^{231}$Pa  & 	968.9  & $^{228}$Ac	\\	
  302.7 &	$^{231}$Pa   & 1001.0	& 	$^{234m}$Pa 	 \\ 		
 323.9 &	$^{223}$Ra & 	1109.5	 & 		$^{211}$Pb 		\\  
   \noalign{\smallskip}\hline
\end{tabular}
\end{table}
It should be mentioned that the observed gamma rays were similar in all three data sets: 180 g, 360 g and 540 g of Zr sample.
The level of trace radioactive impurities present in the Zr is also calculated from the observed activity of the corresponding gamma rays and is given in table~\ref{tab:impurity}. The activity values have been extracted from the 540 g data set, which has largest statistics and the best shielding setup. The activity $A$ is estimated using the eq.~\ref{eg:activity} given below,
\begin{equation}
\label{eg:activity}
{A} =\frac{Y-B}{\eta.~\epsilon.~M}
\end{equation}
where $Y$ and $B$ are the measured counting rates in the signal and background data, respectively. The $\eta$ is the branching fraction of the gamma ray~\cite{toi}, $\epsilon$ is the detection efficiency for the corresponding gamma ray computed with the MC simulations and $M$ is the mass of the sample. The limits of the activity at 90\% C.L. are calculated according to the Feldman-Cousins method~\cite{feldman}. The errors in the estimated activity $A$ include the statistical errors in $Y$ and $B$ as well as the systematic errors in the detection efficiency $\epsilon$. It should be mentioned that $^{235}$U gamma ray lines in natural Zr sample were also reported in an earlier Zr study~\cite{96Zr}. For the $^{235}$U decay chain, the isotopes $^{223}$Ra, $^{227}$Th, $^{211}$Pb activities are not in equilibrium with the $^{235}$U activity level. It can be seen that amongst various decay products of $^{238}$U and $^{232}$Th decay chains, $^{214}$Bi, $^{228}$Ac and $^{208}$Tl show lower yields (and subsequently lower activity). However, the origin of this inconsistency is not clear. It should be also mentioned that the measured levels of $^{238}$U and $^{232}$Th from SIMS analyses are not consistent with that estimated from the spectroscopy studies with the TiLES.

\begin{table}[H]
\centering
\caption{The level of trace radioactive contamination in the Zr sample measured with the TiLES. Limits are given at 90\% C.L. }
\label{tab:impurity}

\begin{tabular}{  lll }
\hline\noalign{\smallskip}
  Source    	& 	$E_{\gamma}$	        &  		Activity  	       \\
          	& 	(keV)        	        & 		(mBq/kg) 	        \\
          	\noalign{\smallskip}\hline\noalign{\smallskip}
$^{210}$Pb	& 	46.5 			& 		12026(3485)		\\
$^{223}$Ra	&	154.2 			& 		3066(167)		\\   
$^{235}$U	&	163.3 			& 		275(62)		\\   
$^{227}$Th	& 	235.9 			& 		2942(192)		\\  
$^{137}$Cs	& 	661.8 			& 		$\leq$ 18.0	        \\
$^{211}$Pb	& 	832.0 			& 		2967(170)		\\
$^{228}$Ac	&	911.2 			& 		56(6)			\\
$^{234m}$Pa	& 	1001.0 			& 		9054(684)		\\
$^{60}$Co	& 	1173.2 			& 		$\leq$4.2 	\\
$^{40}$K	        & 	1460.8 			& 		$\leq$99		\\ 
$^{214}$Bi	& 	1764.5 			& 		27(10)			\\ 
$^{208}$Tl	& 	2614.5 			& 		8.9(0.8)	        \\ 
 	\noalign{\smallskip}\hline
\end{tabular}
\end{table}

\begin{figure}
\centering
\resizebox{0.6\textwidth}{!}{\includegraphics{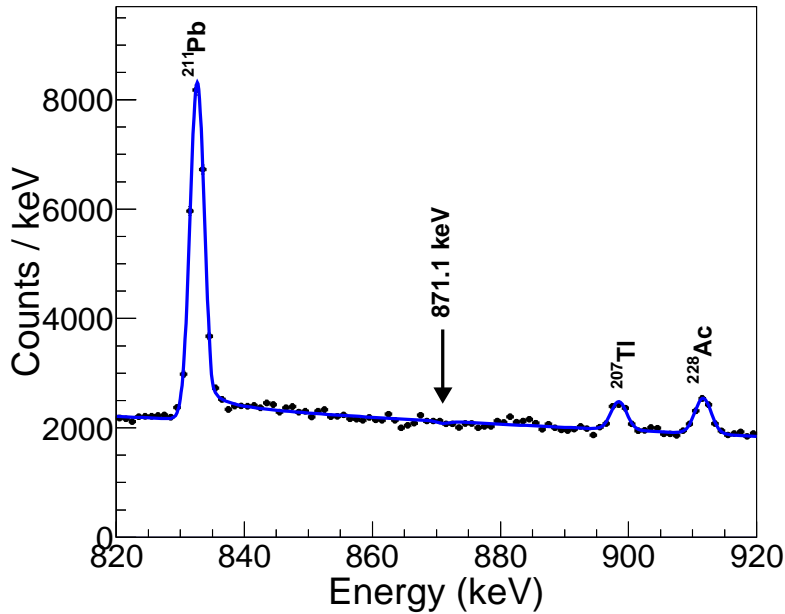}}
\caption{\label{fig:zr540_fit} (Color online) Expanded region ($E_{\gamma}$= 820-920 keV) of the energy spectrum of Zr for the 232 g.y exposure. The fit is shown by the blue line and the black circles represent the data.}
\end{figure}

The energy spectrum in the range 820-920 keV is considered for the calculation of the number of DBD events of interest. The data sets with different Zr masses have been added together after correcting for the calibrations shifts. The energy resolution was verified and was found to be similar for all three data sets. The background is assumed to be linear and is modeled as a first order Chebyshev polynomial in the RooFit analysis framework~\cite{roofit}. The peak at 832.0 keV ($^{211}$Pb) is fitted with a Crystal Ball function due to the presence of an exponential tail on the right side while the 897.8 ($^{207}$Tl) and 911.2 keV ($^{228}$Ac) peaks have been fitted with a Gaussian function. The signal is then introduced at 871.1 keV with a fixed sigma of 1 keV, estimated from the neighbouring 832 keV peak. The 832.0 keV peak is closer to 871.1 keV and has the highest statistics. The number of events under the peak at 871.1 keV is obtained as $-97 \pm 84$. Fig.~\ref{fig:zr540_fit} shows the energy spectrum of Zr in the energy range $E_{\gamma}=$ 820-920 keV for the 232 g.y exposure. The fit is shown by the blue line while the black circles represent the data. The ROI was also varied to 871.1 $\pm$ 50 keV and no significant change was observed in the number of events at 871 keV peak and the background. We have also analysed and fitted three data sets separately. The number of events under the peak at 871.1 keV are similar as compared to the analysis with combined data from the three sets. This is because the shapes of the background and gamma ray peaks are similar in all three data sets.
In the presence of backgrounds, the limit on the half-life $T_{1/2}$ when no signal is observed can be written as
\begin{equation}\label{eq:limit}
{{T}_{1/2}} >\frac{ln2.~N_A.~i}{A}\sum_{k=1}^{3}\frac{M_k.~\epsilon_k.~t_k}{lim~S}
\end{equation}
where $N_A$ is Avogadro's number, {\it{i}} is the isotopic abundance of $^{94}$Zr and $A$ is the molecular mass of Zr. The $\epsilon_k$ is the photopeak detection efficiency of the detector for 871.1 keV, $M_k$ is the mass of the Zr sample and $t_k$ is the counting period for the corresponding data set $k$ (see table~\ref{tab:final}).
As mentioned earlier, the detection efficiency at the gamma ray energy $E_{\gamma}=$ 871.1 keV was determined using the MC detector model~\cite{nima}. The initial kinematics of the gamma rays were generated using the event generator DECAY0~\cite{decay0}. The {$lim~S$} for the number of events obtained from the Gaussian fit at 871.1 keV is calculated as 61 at 90\% C.L. using the Feldman-Cousins procedure~\cite{feldman}. The sensitivity is also determined using the Feldman-Cousins method. The number of events in the ROI is 2000$\times$5 = 10,000 events, which gives the {\it lim $S$} as 184. Thus, the senistivity at 90\% C.L. is three times poorer than the half-life limit obtained with this method.  

\begin{figure}[H]
\centering
\resizebox{0.6\textwidth}{!}{\includegraphics{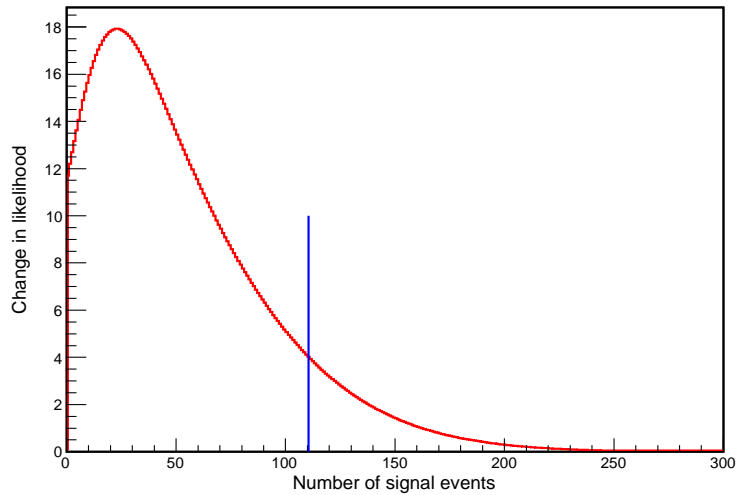}}
\caption{\label{fig:bayesian} (Color online)  Likelihood as a function of number of signal events. The blue line represent the 90\% C.L. limit.}
\end{figure}

An independent estimate of $lim~S$ is also obtained using the Bayesian analysis method~\cite{bayesian}. All components including the peaks and background are floated in the maximum likelihood fit, while the signal width at 871.1 keV is kept fixed at sigma 1 keV.
We obtain an upper limit of 110.5 at 90\% C.L. on the number of signal events $N_{sig}$ ($lim~S$) using eq.~\ref{eq:bayesian} by integrating the likelihood ($\mathcal{L}$) of the fit with fixed values of the $N_{sig}$, as shown in Figure~\ref{fig:bayesian}.
\begin{equation}\label{eq:bayesian}
\int_{0}^{\infty}dN_{sig} = 0.9\times\int_{0}^{\infty}\mathcal{L}(N_{sig})~dN{_{sig}}
\end{equation}
The systematic errors estimated are : 5\% in the MC computed efficiency ($\epsilon_k$), 1\% in the drifts in the energy scale, 1\% in the isotopic composition of $^{94}$Zr and 14\% in the fit parameters of the background model. Hence, the systematic uncertainty in the half-life limit is 16.5\%. 
Systematic uncertainties are included by convoluting the likelihood function with a Gaussian function of width equal to the total uncertainty (16.5\%).  The relevant parameters used in the extraction of the lower limit for the transition are listed in table~\ref{tab:final}. 
The higher value of $lim~S$ is used to quote the final lower half-life limit, i.e., from the Bayesian analysis method.
Using eq.~\ref{eq:limit} and values of the parameters from table~\ref{tab:final}, a lower half-life limit for the DBD of $^{94}$Zr to the $2^{+}_{1}$ excited level of $^{94}$Mo has been established as $T_{1/2} (0\nu + 2\nu) > 3.4\times10^{19}$ y at 90\% C.L. using the total statistics collected with the 232 g.y exposure. The obtained limit corresponds to both 0$\nu\beta\beta$ and 2$\nu\beta\beta$ decay.
The half-life limit obtained at 90\% C.L. in the present work is improved by a factor of $\sim4$ compared to the earlier reported value of 0.8$\times10^{19}$ y~\cite{norman}. The sensitivity is estimated to be $T_{1/2}(0\nu + 2\nu) > 2.0\times10^{19}$ y at 90\% C.L. using the Feldman-Cousins method. 

\begin{table}[H]
\centering
\begin{threeparttable}[b]
\caption{Parameters for the Zr sample.}
\label{tab:final}
\begin{tabular}{  lll }
\hline\noalign{\smallskip}

Parameter	           &	   Symbol	     &	  Value	 		\\
	\noalign{\smallskip}\hline\noalign{\smallskip}
Target Mass ($\rm^{nat}Zr$) &       $M_k$ (g)	     &	(a) 180, (b) 360, (c) 540		\\ 
Efficiency	           & 	$\epsilon_k$ (\%)	     &	 (a) 3.29(0.16), (b) 2.59(0.13), (c) 1.98(0.09) 	\\ 
Live time\tnote{*}	           & 	$t_k$ (days)  	     & 	(a) 15.7, (b) 34.8, (c) 128.3	\\ 
Isotopic Abundance ($\rm^{94}Zr$) &	$i$ (\%)	&	16.58(0.17)	\\
Atomic Mass                 &        $A$ 	     &	91.224(0.002)	\\ 
\noalign{\smallskip}\hline
\end{tabular}

\begin{tablenotes}
\item[*]The time for which the system is active.
\end{tablenotes}
\end{threeparttable}
\end{table}

\section{Summary}
Neutrinoless double beta decay, a rare second-order weak nuclear transition, is of fundamental interest for particle physics. The DBD transitions to the excited states can provide supplementary information for the calculation of NTME for the process. The DBD to excited states can be studied using low background counting facility with HPGe detector having superior energy resolution for gamma rays. In the present work, DBD of $^{94}$Zr to the $2^{+}_{1}$ excited state of $^{94}$Mo at 871.1 keV is studied using a $\sim$ 230 cm$^3$ HPGe detector. No experimental evidence of this decay was found with a 232 g.y exposure of natural Zirconium. The lower limit obtained for the half-life of DBD of $\rm^{94}Zr$ to the $2^{+}_{1}$ excited state of $\rm^{94}Mo$ is $T_{1/2} (0\nu + 2\nu) > 3.4 \times 10^{19}$ y at 90\% C.L using the Bayesian analysis method. The current quoted limit at 90\% C.L. has been improved by a factor of $\sim4$ over the only existing experimental limit. While the sensitivity is estimated to be $T_{1/2} (0\nu + 2\nu) > 2.0\times10^{19}$ y at 90\% C.L. using the Feldman-Cousins method. Measurements with higher sensitivity and different nuclei are required to distinguish between different nuclear models used for NTME calculations.

\section{Acknowledgments}

The authors thank Mr.~M.~S.~Pose, Mr.~K.~V.~Anoop, Mr.~K.~Divekar, Mr.~S.~Mallikarjunachary for assistance during the measurements and Prof.~R.~Palit for useful discussions.

\end{document}